\documentclass[twocolumn,preprintnumbers, amssymb,amsmath,aps,floatfix,nofootinbib,superscriptaddress,showpacs]{revtex4-1}
\usepackage{epsfig}
\usepackage{bm}
\usepackage{amssymb}
\usepackage{amsmath}
\usepackage{color}
\usepackage{subfigure}
\usepackage[colorlinks,
            linkcolor=blue,
            anchorcolor=black,
            citecolor=blue
            ]{hyperref}
\usepackage{slashed}
\usepackage{soul}
\usepackage{xcolor}
\definecolor{darkred}{rgb}{0.55, 0, 0}

\newcommand{\eg}{\textit{e.g.}}  

\newcommand{\ie}{\textit{i.e.}}

\begin{document}


\title{Unified description of Sivers and Boer–Mulders asymmetries from twist-3 correlations} 

\author{Zhimin Zhu}
\email{zhuzhimin@impcas.ac.cn}
\affiliation{State Key Laboratory of Heavy Ion Science and Technology, Institute of Modern Physics, Chinese Academy of Sciences, Lanzhou, Gansu, 730000, China}
\affiliation{School of Nuclear Science and Technology, University of Chinese Academy of Sciences, Beijing, 100049, China}

\author{Jiangshan Lan}
\email{Jiangshanlan@impcas.ac.cn}
\affiliation{State Key Laboratory of Heavy Ion Science and Technology, Institute of Modern Physics, Chinese Academy of Sciences, Lanzhou, Gansu, 730000, China}
\affiliation{School of Nuclear Science and Technology, University of Chinese Academy of Sciences, Beijing, 100049, China}

\author{Chandan Mondal}
\email{mondal@impcas.ac.cn}
\affiliation{State Key Laboratory of Heavy Ion Science and Technology, Institute of Modern Physics, Chinese Academy of Sciences, Lanzhou, Gansu, 730000, China}
\affiliation{School of Nuclear Science and Technology, University of Chinese Academy of Sciences, Beijing, 100049, China}

\author{Xingbo Zhao}
\email{xingbozhao@impcas.ac.cn}
\affiliation{State Key Laboratory of Heavy Ion Science and Technology, Institute of Modern Physics, Chinese Academy of Sciences, Lanzhou, Gansu, 730000, China}
\affiliation{School of Nuclear Science and Technology, University of Chinese Academy of Sciences, Beijing, 100049, China}

\author{James P. Vary}
\email{jvary@iastate.edu}
\affiliation{Department of Physics and Astronomy, Iowa State University, Ames, IA 50011, USA}

\collaboration{BLFQ Collaboration}

\begin{abstract}
  We present the first calculation of the Efremov-Teryaev-Qiu-Sterman functions and associated twist-3 quark-gluon correlation functions for both the proton and pion. 
  These functions are determined using the light-front wave functions obtained by diagonalizing a light-front effective Hamiltonian within a Fock space truncated to include a dynamical gluon. 
  We compute the twist-3 correlations in the hard-pole region and extrapolate them to the soft-gluon pole limit.
  After the scale evolutions, our predictions demonstrate quantitative consistency with recent experimental extractions, providing a unified description of the Sivers and Boer-Mulders asymmetries from a light-front Hamiltonian approach.
\end{abstract}
\maketitle

{\it Introduction.}---%
Understanding the internal structure of hadrons, particularly the correlations between parton (quarks and gluons) transverse momentum and hadron spin, remains a central frontier of quantum chromodynamics (QCD)~\cite{Diehl:2003ny,Accardi:2012qut,Gross:2022hyw,Boussarie:2023izj}, the well-established theory of strong interactions.
These correlations manifest experimentally as spin-dependent asymmetries, such as the Sivers asymmetry~\cite{HERMES:2009lmz,COMPASS:2008isr,JeffersonLabHallA:2011ayy,JeffersonLabHallA:2014yxb,COMPASS:2016led} and the Boer–Mulders effect~\cite{NuSea:2006gvb,NuSea:2008ndg,Longo:2019bih}, observed in semi-inclusive deep inelastic scattering (SIDIS) and Drell–Yan processes.
Theoretically, these phenomena can be described within two complementary pictures: transverse momentum dependent factorization~\cite{Ji:2004xq,Ji:2004wu,Collins:2011zzd}, and twist-3 collinear factorization involving quark-gluon correlations~\cite{Qiu:1991wg,Qiu:1991pp,Ji:2006ub,Ji:2006vf,Ji:2006br}.

Central to this understanding are the time-reversal–odd (``T-odd'') parton distributions, such as the Sivers function $f_{1T}^\perp$ and the Boer–Mulders function $h_1^\perp$~\cite{Sivers:1990fh,Sivers:1989cc,Boer:1997nt}.
At the operator level, these distributions are related to quark–gluon–quark correlation functions in collinear twist-3 factorization: the Efremov–Teryaev–Qiu–Sterman (ETQS) function $T_F(x,x)$ corresponds to the first transverse moment of the Sivers function, $f_{1T}^{\perp(1)}(x)$, while its chiral-odd counterpart, $T_F^{(\sigma)}(x,x)$, is associated with the first transverse moment of the Boer–Mulders function, $h_{1}^{\perp(1)}(x)$~\cite{Boer:2003cm,Ma:2003ut}.
Physically, these twist-3 functions defined at the soft-gluon pole (SGP) limit ($x_1=x_2=x$, making $x_g = |x_1 - x_2| = 0$) represent the quantum interference between an active quark and a quark accompanied by a soft gluon~\cite{Qiu:1991pp,Qiu:1991wg,Qiu:1998ia}. 
More generally, these correlations can be extended to the hard pole region ($x_1 \neq x_2$) involving hard gluons~\cite{Qiu:1991wg,Ji:2006ub,Ji:2006vf,Zhou:2009jm}.

Despite their importance, first-principles calculations of twist-3 correlations remain a significant challenge.
Phenomenological extractions have provided valuable constraints on ETQS functions, but they depend on model assumptions and choices of factorization scheme~\cite{Cammarota:2020qcw,Echevarria:2020hpy,Bacchetta:2020gko}.
Lattice QCD has begun to access moments of transverse-momentum-dependent distributions (TMDs) and twist-3 observables, but direct calculations of ETQS functions in hadrons such as the proton and pion remain at an early stage~\cite{Musch:2011er,Lin:2017snn}.

In this Letter, we present the first simultaneous calculation of the ETQS functions and the associated twist-3 functions for both the proton and pion within the Basis Light-Front Quantization (BLFQ) framework~\cite{Vary:2009gt}.
BLFQ is a non-perturbative Hamiltonian formalism that solves for the light-front wave functions (LFWFs) of hadrons by diagonalizing the light-front QCD Hamiltonian in a truncated Fock space. 
It has been successfully applied to a wide range of hadron-structure observables~\cite{Vary:2025yqo,Lan:2019vui,Lan:2019img,Lan:2019rba}.
By truncating the Fock space to include the valence sector plus one dynamical gluon, \ie, $|qqqg\rangle$ for the proton and $|q\bar{q}g\rangle$ for the pion~\cite{Xu:2021wwj,Lan:2021wok}, we compute the multiparton matrix elements and extract the twist-3 correlations to investigate the Sivers and Boer–Mulders effects.
After applying QCD evolution, our results provide a unified description of the spin-flavor structure of light hadrons, offering new insights into the non-perturbative origin of spin asymmetries.

{\it Pion and proton LFWFs from BLFQ.}---%
We evaluate the LFWFs of the pion and proton by diagonalizing the light-front effective Hamiltonian, $H_{\rm{eff}} \equiv P^+ (P^-_{\rm{QCD}} + P_{\rm{C}}^-)$, within the BLFQ framework~\cite{Vary:2009gt,Vary:2025yqo}.
Following Refs.~\cite{Xu:2021wwj,Lan:2021wok}, the hadron states are truncated to the Fock sectors containing valence quarks and up to one dynamical gluon,
\begin{equation}
\begin{aligned}
  |\text{pion}\rangle &= \Psi_{\pi,2} | q\bar{q} \rangle + \Psi_{\pi,3} | q\bar{q}g \rangle, \\
  |\text{proton}\rangle &= \Psi_{\rm{p},3} | qqq \rangle + \Psi_{\rm{p},4} | qqqg \rangle.
\end{aligned}
\end{equation}
In the light-front gauge $A^+ = 0$, the QCD Hamiltonian $P_{\rm QCD}^-$, with non-vanishing contributions in our chosen sectors, includes the kinetic energy of quarks and gluons, the quark-gluon interaction, and the instantaneous gluon exchange interaction~\cite{Brodsky:1997de}. 
Explicitly, the kinetic terms include the bare quark mass $m_0$ and a phenomenological gluon mass $m_g$~\cite{Cornwall:1981zr,Alkofer:2000wg,Deur:2016tte}. 
The interaction terms involve the coupling constant $g$.
Following the Fock sector dependent renormalization procedure~\cite{Karmanov:2008br}, renormalization effects are incorporated by introducing a counterterm $\delta m_q = m_0 - m_q$, where $m_q$ is the renormalized quark mass.
Furthermore, we distinguish the kinetic mass from the vertex mass $m_f$ in the interaction terms to compensate for missing non-perturbative contributions due to Fock-space truncation~\cite{Glazek:1992aq}.

Confinement is introduced via $P^-_\mathrm{C}$ in the valence sector. The longitudinal and transverse confining potential read as~\cite{Li:2015zda}
\begin{align}
    P^+P^-_{\mathrm{C}}=\frac{\kappa^4}{2}\sum_{i\neq j}\left[\vec{r}_{ij\perp}^2-\frac{\partial_{x_i}(x_ix_j\partial_{x_j})}{(m_i+m_j)^2}\right],
    \label{eq:confinement}
\end{align}
where $\vec{r}_{ij\perp}=\sqrt{x_ix_j}(\vec{r}_{i\perp}-\vec{r}_{j\perp})$ is the relative coordinate, $\partial_{x}\equiv (\partial/\partial x)_{r_{ij\perp}}$, and $\kappa$ is the confinement strength.

The Hamiltonian is diagonalized in a basis $|\alpha\rangle=\otimes_i|k_i, n_i, m_i, \lambda_i\rangle$, where longitudinal momenta are discretized as $p_i^+ = 2\pi k_i / L$ in a box of length $2L$.
The antiperiodic (periodic) boundary conditions for fermions (bosons) result in half-integer (integer) dimensionless momenta $k_i$.
The transverse motions are expanded in terms of two-dimensional harmonic oscillator functions $\Phi_{n_im_i}(\vec{p}_{\perp i};b)$ with a scale parameter $b$. $\lambda_i$ denotes the helicity of the $i$-th parton.
Note that the color indices of the Fock states are omitted for simplicity.
Ultraviolet and infrared regularizations are provided by the longitudinal resolution $K = \sum_i k_i$ and the transverse cutoff $N_{\max} \ge \sum_i(2n_i+|m_i|+1)$~\cite{Zhao:2014xaa}.
The resulting LFWF for an $\mathcal{N}$-particle sector can be expressed schematically as
\begin{align}
    \Psi_{\mathcal{N}}^{\Lambda}({\{x_{i}, \vec{p}_{i \perp}, \lambda_{i} \}})
    =\sum_{\{n_i,m_i\}}\psi^{\Lambda}_{\mathcal{N}}(\{\alpha_i\}) \prod_{i=1}^{\mathcal{N}} \Phi_{n_{i}m_i}\left(\vec{p}_{i \perp} ; b\right),\nonumber 
\end{align}
where $\psi^{\Lambda}_{\mathcal{N}}$ are the eigenvectors obtained from diagonalization of the Hamiltonian $H_{\rm{eff}}$ in the above discretized basis.
For the spinless pion, we omit the helicity label in its LFWFs and eigenvectors.

To investigate the ETQS functions and associated twist-3 correlations, we employ LFWFs determined using the parameter sets established in Refs.~\cite{Lan:2021wok,Xu:2021wwj}, 
which have successfully described the electromagnetic form factors, charge radii, and parton distributions of the pion and proton~\cite{Lan:2021wok,Xu:2021wwj,Zhu:2023lst,Lan:2024ais,Wu:2024hre,Yu:2024mxo,Zhu:2024awq,Zhang:2025nll}.

{\it Twist-3 Correlations.}---%
The twist-3 quark-gluon correlations, particularly the ETQS functions~\cite{Efremov:1981sh,Qiu:1991pp}, are fundamental to understanding the spin asymmetries in hard scattering processes~\cite{Qiu:1991pp,Zhou:2008fb}. 
These correlations are encoded in the multiparton correlation matrix element. 
For a hadron with momentum $P$ and spin $S$, the correlator is defined as~\cite{Ellis:1982cd,Ellis:1982wd,Zhou:2008mz,Zhou:2009jm}
\begin{align}
  \hat{M}_{F,\alpha\beta}^\mu&(x_1, x_2) = 
  \int \frac{\mathrm{d} y_1^{-} \mathrm{d} y_2^{-}}{(2\pi)^2} e^{i\frac{y_1^-}{2}(x_1+x_2)P^+}e^{i y_2^{-}\left(x_2-x_1\right) P^{+}} \nonumber\\
  &\times \langle P, S|\bar{\psi}_\beta(-\tfrac{y_1^{-}}{2}) g F^{+\mu}\left(y_2^{-}\right) \psi_\alpha(\tfrac{y_1^{-}}{2})| P, S\rangle,
  \label{eq:MF_def}
\end{align}
where $F^{+\mu}$ is the gluon field strength tensor with transverse index $\mu=1,2$, and the suppressed gauge links are omitted for brevity.
The correlator for the proton is parametrized into four twist-3 functions~\cite{Zhou:2008mz,Zhou:2009jm},
\begin{align}
  \hat{M}_F^\mu (x_1, x_2) =\frac{M}{2}\Big[ &T_F \epsilon_{\perp}^{\nu \mu} S_{\perp \nu} \gamma^- +\tilde{T}_F i S_{\perp}^\mu \gamma_5 \gamma^- \nonumber\\
  + & \tilde{T}_F^{(\sigma)} i {\Lambda} \gamma_5 \gamma_{\perp}^\mu \gamma^- + T_F^{(\sigma)} i \gamma_{\perp}^\mu \gamma^- \Big],
  \label{eq:proton_param}
\end{align}
where the arguments $(x_1,x_2)$ are implicit, $M$ is the hadron mass, ${\Lambda} $ is the helicity,  and $S_\perp$ is the transverse spin vector.
The parity and time-reversal invariance indictates that $T_F(x_1,x_2)$ and $T_F^{(\sigma)}(x_1,x_2)$ are symmetric under $x_1 \leftrightarrow x_2$, while $\tilde{T}_F(x_1,x_2)$ and $\tilde{T}_F^{(\sigma)}(x_1,x_2)$ are antisymmetric~\cite{Qiu:1991wg,Zhou:2009jm}.
For the spinless pion, the spin-dependent terms vanish, and only the chiral-odd twist-3 correlation function $T_F^{(\sigma)}$ survives,
\begin{align}
  \hat{M}_{F}^\mu(x_1, x_2) = M \frac{i}{2} T_F^{(\sigma)} (x_1, x_2) \gamma_{\perp}^\mu \gamma^-.
  \label{eq:ETQS_pion}
\end{align}

In the ``diagonal'' limit ($x_1=x_2=x$), corresponding to the SGP region, the ETQS function $T_F(x,x)$ and the chiral-odd counterpart $T_F^{(\sigma)}(x,x)$ satisfy, from the leading order derivation, the relations with the first transverse moments of the T-odd TMDs~\cite{Boer:2003cm,Ma:2003ut,Zhou:2008fb,Zhou:2009jm,Kang:2012em,Wang:2018naw}\footnote{
  The T-odd TMDs are process-dependent due to the gauge-link structure. 
  Their signs reverse between SIDIS and Drell-Yan processes~\cite{Collins:2002kn}. 
  We present all results with the Drell-Yan definition. 
}. 
Specifically, 
\begin{align}
  \pi T_{F}(x,x) &= \int \mathrm{d}^2k_\perp \frac{k_\perp^2}{2M^2} f_{1T}^\perp (x, k_\perp^2) \equiv f_{1T}^{\perp(1)} (x), \label{eq:QS_Sivers}\\
  \pi {T}_F^{(\sigma)}(x,x) &= \int \mathrm{d}^2k_\perp \frac{k_\perp^2}{2M^2} h_{1}^\perp (x, k_\perp^2) \equiv h_{1}^{\perp(1)} (x). \label{eq:QS_BM}
\end{align}
While perturbative matching also relates these twist-3 functions to the large-$k_\perp$ tail of these TMDs~\cite{Zhou:2009jm,Sun:2013hua}, our focus here remains on the non-perturbative SGP correlations.

{\it Numerical results.}---%
We determine the LFWFs of pions and protons by solving the light-front stationary Schr\"odinger equation. 
The resulting spectra yield ground-state masses of $M_{\rm{p}} = 0.956\;\mathrm{GeV}$ and $M_\pi = 0.139\;\mathrm{GeV}$.
Using these LFWFs, we compute the matrix elements of the operator in Eq.~(\ref{eq:MF_def}). 
For the proton, the four independent ETQS functions are isolated by projecting the matrix element onto the basis of helicity amplitudes.
For the pion, the function is extracted via the trace projection $T_F^{(\sigma)} =\frac{1}{2M} \mathrm{Tr} [ \hat{M}^{\mu}_{F} \sigma^{+\mu} ]$.
Technical details of the numerical procedure are provided in the Supplemental Material.

The LFWFs derived from the effective Hamiltonian define the distributions at an initial model scale, $\mu_0$.
To compare our predictions with experimental data measured at higher energy scales, QCD evolution is essential.
While the evolution equations for twist-3 correlators have been studied~\cite{Zhou:2008mz,Kang:2008ey,Braun:2009mi,Vogelsang:2009pj,Kang:2012em,Echevarria:2020hpy}, a comprehensive numerical implementation remains challenging due to the mixing of different correlation functions and the covering of the full kinematic domain, including the hard-pole and SGP regions~\cite{Kang:2012em}.
In this work, we adopt a widely used approximate evolution scheme~\cite{Sun:2013hua,Wang:2017onm,Bacchetta:2020gko,Kou:2023ady,Cheng:2024gyv}. 
For the ETQS function, $T_F(x,x)$ is evolved using the unpolarized splitting kernel $P_{qq}$
~\cite{Sun:2013hua,Cammarota:2020qcw,Bacchetta:2020gko},
and for the chiral-odd twist-3 function, $T_F^{(\sigma)}(x,x)$ is evolved using the kernel transversity splitting kernel $\Delta_T P_{qq}$~\cite{Pasquini:2014ppa,Cheng:2024gyv}.
The evolution functions are degenerated to the DGLAP equations after these approximations.

Explicitly, we evolve our results from the initial scale to the relevant experimental scales of $\mu^2 = 4\;\mathrm{GeV}^2$ and $\mu^2 = 100\;\mathrm{GeV}^2$ by numerically solving the leading-order DGLAP equations. 
Following Refs.~\cite{Lan:2021wok,Xu:2021wwj}, the initial scales, {$\mu^2_{0,\mathrm{p}} = 0.079\;\mathrm{GeV}^2$} for the proton and $\mu^2_{0,\pi} = 0.222\;\mathrm{GeV}^2$ for the pion, are determined by matching moments from a global QCD analysis. 
For the proton, we match the combined valence quark moment $\langle x\rangle_u + \langle x\rangle_d = 0.3742$ at $\mu^2 = 10\;\rm{GeV}^2$~\cite{deTeramond:2018ecg}. 
For the pion, we match the total first moment of the valence distribution, $\langle x \rangle_{\rm{valence}} = 0.480$, at $\mu^2 = 5\;\rm{GeV}^2$~\cite{Barry:2018ort}. 
To account for uncertainties of the twist-3 distributions, we estimate error bands by varying both the initial scale $\mu_0^2$ and the coupling constant $g^2$ by $\pm 10\%$.

\begin{figure}[h]
  \centering
  \includegraphics[width=0.99\columnwidth]{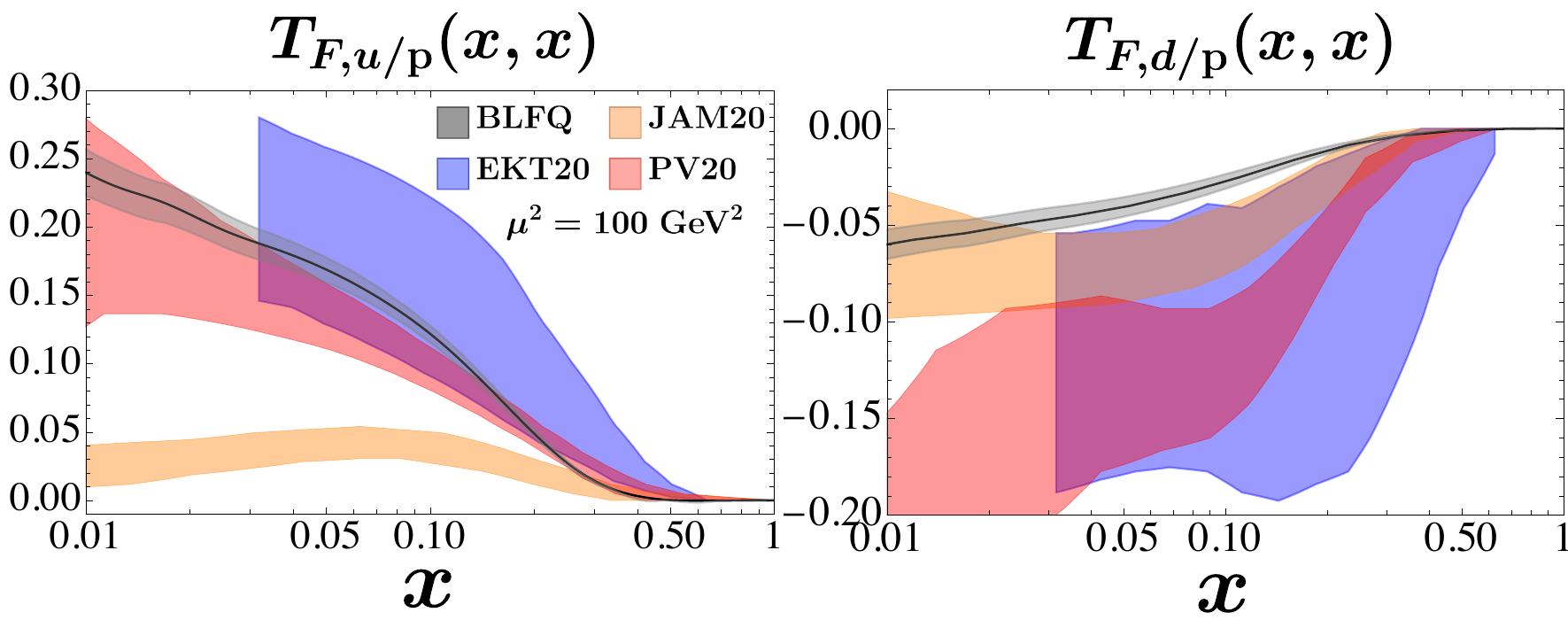}

    \caption{
      The proton ETQS function $T_F(x,x)$ for $u$ (left) and $d$ (right) quarks at $\mu^2 = 100\; \mathrm{GeV}^2$.
      The black curves with bands represent the BLFQ results with uncertainty estimates.
      Color bands represent experimental extractions from JAM20~\cite{Cammarota:2020qcw}, EKT20~\cite{Echevarria:2020hpy}, and PV20~\cite{Bacchetta:2020gko}.
    }
    \label{fig:TF_p}
\end{figure}
Figure~\ref{fig:TF_p} presents the BLFQ prediction for the proton ETQS function $T_F(x,x)$ at $\mu^2 = 100\; \mathrm{GeV}^2$.
Our results (black curves) show a clear sign difference between the $u$ and $d$ quarks, consistent with the experimental extractions from JAM20~\cite{Cammarota:2020qcw}, EKT20~\cite{Echevarria:2020hpy}, and PV20~\cite{Bacchetta:2020gko}.
The magnitude of the $u$-quark distribution is notably larger than that of the $d$-quark, reinforcing the flavor asymmetry observed in T-odd effects.

\begin{figure}[t]
  \centering
    \includegraphics[width=0.99\columnwidth]{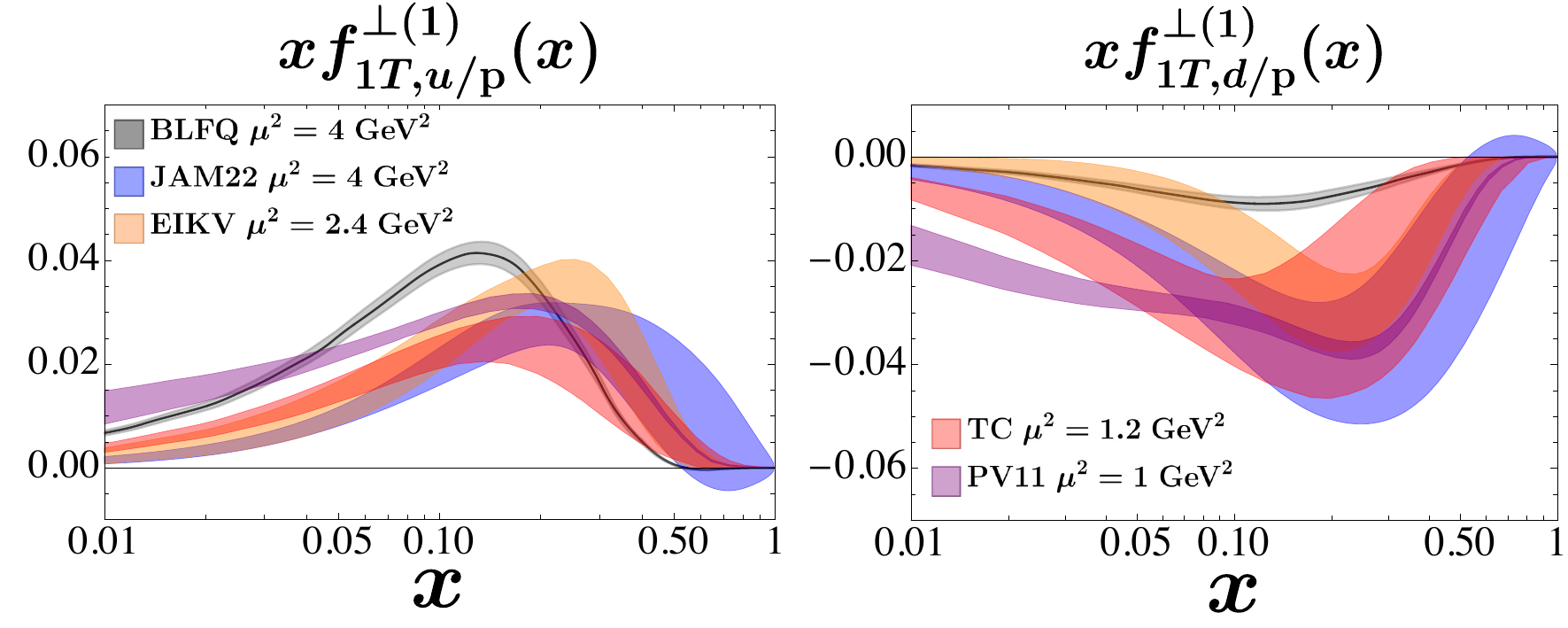}
    \caption{ 
      The first transverse moment of the proton Sivers function, $x f_{1T}^{\perp(1)}(x)$.
      BLFQ predictions (black curves) are compared with extractions from JAM22~\cite{Gamberg:2022kdb}, EIKV~\cite{Echevarria:2014xaa}, TC~\cite{Boglione:2018dqd}, and PV11~\cite{Bacchetta:2011gx}.
    }
    \label{fig:f1Tperp}
\end{figure}
In Fig.~\ref{fig:f1Tperp}, we compare the first transverse moment of the Sivers function, $x f_{1T}^{\perp(1)}(x)$, which is directly related to $T_F(x,x)$ via the model-independent relation in Eq.~(\ref{eq:QS_BM}).
The BLFQ results at $\mu^2 = 4 \;\mathrm{GeV}^2$ are compared with global fits from JAM22~\cite{Gamberg:2022kdb}, EIKV~\cite{Echevarria:2014xaa}, TC~\cite{Boglione:2018dqd}, and PV11~\cite{Bacchetta:2011gx} at different scales.
A direct comparison is more difficult because the evolution of TMDs is achieved in a different framework.
Nevertheless, the agreement in sign and shape confirms that the LFWFs correctly capture the spin-orbit correlations required to generate the Sivers asymmetry.

Although there is a series of the Boer-Mulders functions in phenomenological calculations~\cite{Gamberg:2007wm,Courtoy:2008dn,Courtoy:2009pc,Barone:2009hw,Gamberg:2009uk,Pasquini:2010af,Lu:2012hh,Pasquini:2014ppa,Wang:2017onm,Li:2019uhj,Tan:2022kgj,Gurjar:2023uho} and experimental extractions~\cite{Lu:2009ip,Longo:2019bih}, we do not compare them with other results due to the lack of consideration of evolution effects or different evolution frameworks.
Instead, we will compare our BLFQ results between the proton and pion.

\begin{figure}[h]
  \centering
      \includegraphics[width=0.99\columnwidth]{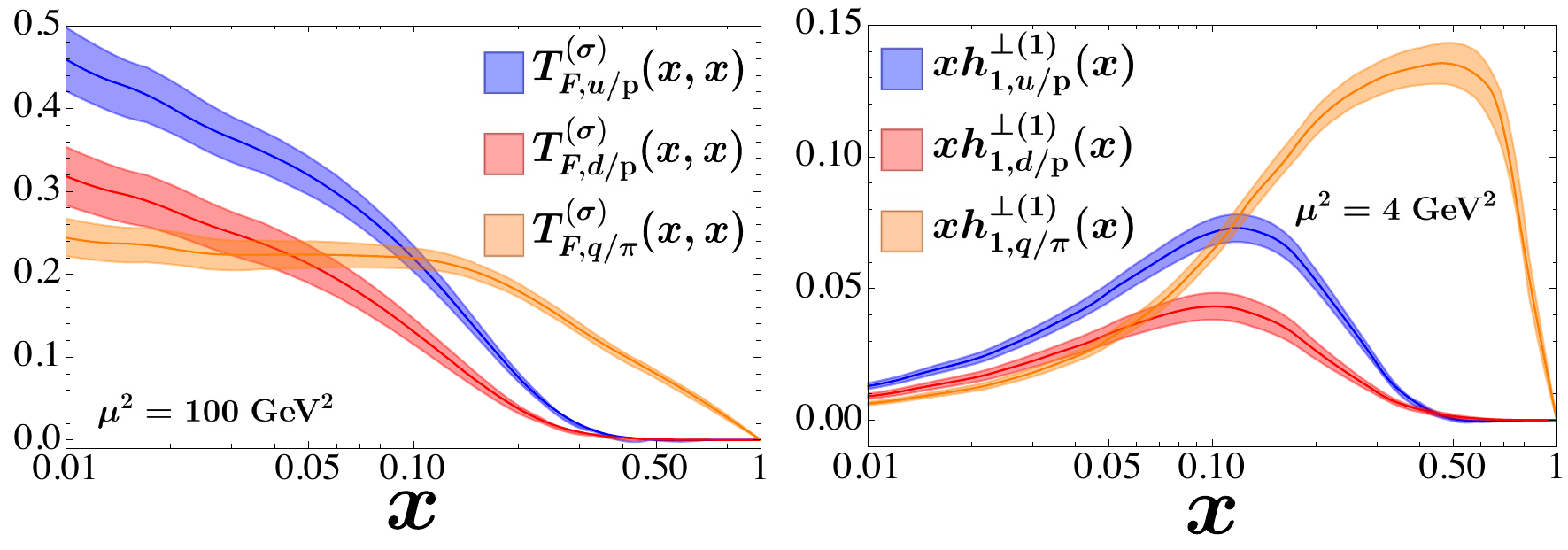}
    \caption{
      Comparison of $T_{F}^{(\sigma)}(x,x)$ and $xh_{1}^{\perp(1)}(x)$ for the proton and pion.
      Left: The associated twist-3 function $T_{F}^{(\sigma)}(x,x)$ at $\mu^2 = 100\; \mathrm{GeV}^2$.
      Right: The first transverse moment of the Boer-Mulders function $x h_{1}^{\perp(1)}(x)$ at $\mu^2 = 4\; \mathrm{GeV}^2$.
      Blue and red curves denote the $u$ and $d$ quarks in the proton, respectively; orange curves denote the light quark in the pion.
      }
    \label{fig:TFsigma_comparison}
\end{figure}
Figure~\ref{fig:TFsigma_comparison} shows the results for $T_F^{(\sigma)}(x,x)$ and $xh_{1}^{\perp(1)}(x)$.
The left panel shows the associated twist-3 function $T_{F}^{(\sigma)}(x,x)$ at $\mu^2 = 100 \;\mathrm{GeV}^2$.
The proton and pion results share the same sign.
The right panel shows the corresponding first transverse moment of the Boer-Mulders function, $x h_{1}^{\perp(1)}(x)$.
The consistent sign between the proton and pion Boer-Mulders functions aligns with phenomenological calculations and experimental extractions.
Besides, their relative are also consistent with phenomenological calculations and experimental extractions.

{\it Conclusion.}---%
This work presents the first calculation of the twist-3 quark-gluon correlation functions for both the proton and the pion using the BLFQ framework.
By incorporating a dynamical gluon into the LFWFs, we successfully capture the essential quantum interference effects for generating the T-odd spin asymmetries.
Our results for the proton ETQS function $T_F(x,x)$ exhibit a flavor dependence and magnitude consistent with recent phenomenological extractions.
In addition, we predict the chiral-odd function $T_F^{(\sigma)}(x,x)$ in both the proton and pion.
These two types of functions correspond to the first moments of the Sivers and Boer–Mulders functions, respectively.
The quantitative agreement achieved after QCD evolution suggests that the truncated Fock space, $|qqqg\rangle$ for the proton and $|q\bar{q}g\rangle$ for the pion, effectively encodes the dominant nonperturbative dynamics governing these higher-twist observables.
Overall, this work establishes a direct connection between the fundamental light-front wave functions and high-energy spin-dependent observables, providing a powerful theoretical framework for exploring the internal structure of hadrons.

\noindent
{\bf Acknowledgments:} 
We thank Hongxi Xing for useful discussions.
This work is supported by the National Natural Science Foundation of China under Grant No.~12375143 and No.~12305095, by the Gansu International Collaboration and Talents Recruitment Base of Particle Physics (2023-2027), by the Senior Scientist Program funded by Gansu Province, Grant No.~25RCKA008.
Z. Zhu is supported by China Association for Science and Technology.
J. Lan is supported by the Special Research Assistant Funding Project, Chinese Academy of Sciences, by Gansu Provincial Young Talents Program,  and by the Natural Science Foundation of Gansu Province, China, Grant No.~23JRRA631.
X. Zhao is supported by Key Research Program of Frontier Sciences, Chinese Academy of Sciences, Grant No.~ZDBS-LY-7020, by the Foundation for Key Talents of Gansu Province, by the Central Funds Guiding the Local Science and Technology Development of Gansu Province, Grant No.~22ZY1QA006, by international partnership program of the Chinese Academy of Sciences, Grant No.~016GJHZ2022103FN, and by the Strategic Priority Research Program of the Chinese Academy of Sciences, Grant No.~XDB34000000.
J. P. Vary is supported by the Department of Energy under Grant No.~DE-SC0023692.  A portion of the computational resources were also provided by Taiyuan Advanced Computing Center.

\bibliographystyle{apsrev4-1}
\bibliography{ref.bib}

\section*{SUPPLEMENTAL MATERIAL}\label{supp}

In this Supplemental Material, we provide detailed derivations of the LFWF overlap representations for the twist-3 correlations and present the numerical results at the initial scale.

\section{Overlap Representations}
\subsection{Proton}
For spin-1/2 baryons, the state $|P, S\rangle$ with momentum $P$ and spin $S$ is expanded in terms of the light-front helicity eigenstates with helicity $\Lambda$.
The twist-3 quark-gluon correlation functions are extracted from the matrix elements defined in Eq.~(\ref{eq:MF_def}) of the main text. 
We define the helicity-dependent matrix elements projected by a generic Gamma matrix $\Gamma$ as,
\begin{align}
  &\hat{M}^{\mu[\Gamma]}_{F,\Lambda \Lambda^\prime}(x_1,x_2)=
  \int \frac{\mathrm{d} y_1^{-} \mathrm{d} y_2^{-}}{(2\pi)^2} e^{i\frac{y_1^-}{2}(x_1+x_2)P^+}e^{i y_2^{-}\left(x_2-x_1\right) P^{+}} \nonumber\\
  &\times \langle P, \Lambda|\bar{\psi}(-\tfrac{y_1^{-}}{2})\Gamma g F^{+\mu}\left(y_2^{-}\right) \psi(\tfrac{y_1^{-}}{2})| P, \Lambda^\prime \rangle.
  \label{eq:M_gamma_def}
\end{align}

By selecting specific Dirac matrices $\Gamma$, we invert the parameterization to isolate the four independent twist-3 functions for the proton, as follows:
\begin{align}
  \tilde{T}_F^{(\sigma)}(x_1,x_2)& = \frac{i}{4 M}\Big[\hat{M}^{1[i \sigma^{1+} \gamma^5]}_{F,++} - \hat{M}^{1[i \sigma^{1+} \gamma^5 ]}_{F,--}\Big], \label{eq:proj_1}\\
  T_F^{(\sigma)} (x_1,x_2) &= \frac{i}{4 M}\Big[\hat{M}^{1[i\sigma^{1+}]}_{F,++} + \hat{M}^{1[i\sigma^{1+}]}_{F,--}\Big], \label{eq:proj_2}\\
  T_{F}(x_1,x_2)& = \frac{i}{4 M}\Big[\hat{M}^{1[\gamma^+]}_{F,+-} - \hat{M}^{1[\gamma^+]}_{F,-+}\Big], \label{eq:proj_3}\\
  \tilde{T}_F(x_1,x_2) & = -\frac{i}{4 M}\Big[ \hat{M}^{1[\gamma^+ \gamma^5 ]}_{F,+-} + \hat{M}^{1[\gamma^+ \gamma^5 ]}_{F,-+}\Big]. \label{eq:proj_4}
\end{align}
Here, the subscripts $\Lambda \Lambda^\prime =++, --, +-, -+$ denote the proton helicities, and we select the transverse index $\mu=1$ for explicit calculation.

We derive the overlap representation by substituting the mode expansion of the quark and gluon field operators (following BLFQ conventions~\cite{Wiecki:2014ola}) and the Fock state expansion into Eq.~(\ref{eq:M_gamma_def}).
Crucially, the operator $g F^{+\mu}$ couples the $\mathcal{N}$-particle sector to the $(\mathcal{N}+1)$-particle sector (\eg, $|uud\rangle$ and $|uudg\rangle$).
The resulting explicit overlap formula for the proton is given by:
\begin{align}
  &\hat{M}_{F,\Lambda\Lambda^\prime }^{\mu[\Gamma]}\left({x}_1, {x}_2\right)  = - 
  \frac{gC_F{K}^{3/2}}{16 \pi^{13/2}}\nonumber\\
  &\times\sum_{\{\lambda_i\}}
  \int 
  \mathrm{d}\xi_1\mathrm{d}^2 p_{1\perp} \mathrm{d}\xi_2\mathrm{d}^2 p_{2\perp} \mathrm{d}\xi_1^\prime\mathrm{d}^2 p_{1\perp}^{\prime}
  \nonumber\\
  &\times
  \Big[\frac{1}{\sqrt{\xi_4^\prime}}
  \Psi^{\Lambda*}_{\mathrm{p},3}(1,2,3)
  \Psi^{\Lambda^\prime}_{\mathrm{p},4}(1^\prime,2,3,4^\prime)
  \bar{u}(1)i\Gamma u(1^\prime)
  \epsilon^{\mu}(4^\prime)
  \nonumber\\
  &\quad \times
  \delta(x_2-\xi_1)\delta(x_1-\xi_1^\prime)
  + (x_1 \leftrightarrow x_2, \Lambda \leftrightarrow \Lambda^\prime)^*\Big].
  \label{eq:amplitudes_proton}
\end{align}
In this expression, we employ the shorthand notation $j \equiv (\xi_j, \vec{p}_{j\perp }, \lambda_j)$ for the parton variables. 
The indices $1, 2, 3$ refer to the active quark and two spectators in the $|uud\rangle$ sector, while $1^\prime, 2, 3, 4^\prime$ refer to the active quark, spectators, and gluon in the $|uudg\rangle$ sector, respectively.
Momentum conservation is imposed in each Fock sector, $\sum_i \xi_i = 1$ and $\sum_i \vec{p}_{i\perp} = 0$.
The spinors $\bar{u}, u$ and the gluon polarization vector $\epsilon^\mu$ follow standard BLFQ conventions~\cite{Wiecki:2014ola}.
${K}$ is the longitudinal resolution associated with the longitudinal truncation, $C_F$ is the color factor, and $x_{1,2}$ are the arguments of the twist-3 correlation.

Employing the amplitudes in Eq.~(\ref{eq:amplitudes_proton}) and the projections in Eqs.~(\ref{eq:proj_1})-(\ref{eq:proj_4}), we obtain the ETQS function and associated twist-3 functions of the proton at the initial scale.

\subsection{Pion}
For the pion, the procedure is analogous, considering the interference between $|q\bar{q}\rangle$ and $|q\bar{q}g\rangle$. The overlap representation of the twist-3 matrix element is
\begin{align}
  &\hat{M}_{F }^{\mu[\Gamma]}\left({x}_1, {x}_2\right)  = - 
  \frac{gC_F{K}^{3/2}}{4 \pi^{9/2}}
  \sum_{\{\lambda_i\}}
  \int 
  \mathrm{d}\xi_1\mathrm{d}^2 p_{1\perp} \mathrm{d}\xi_1^\prime\mathrm{d}^2 p_{1\perp}^{\prime}
  \nonumber\\
  &\times
  \Big[\frac{1}{\sqrt{\xi_3^\prime}}
  \Psi^{*}_{\pi,2}(1,2)
  \Psi^{}_{\pi,3}(1^\prime,2,3^\prime)
  \bar{u}(1)i\Gamma u(1^\prime)
  \epsilon^{\mu}(3^\prime)
  \nonumber\\
  &\quad \times
  \delta(x_2-\xi_1)\delta(x_1-\xi_1^\prime)
  + (x_1 \leftrightarrow x_2)^*\Big].
  \label{eq:amplitudes_pion}
\end{align}
The indices $1, 2$ refer to the active quark and the spectator antiquark in the $|q\bar{q}\rangle$ sector, while $1^\prime, 2, 3^\prime$ refer to the active quark, spectator, and gluon in the $|q\bar{q}g\rangle$ sector, respectively.
Note that for the pion, the twist-3 function is extracted directly via the trace projection given in the main text, utilizing Eq.~(\ref{eq:amplitudes_pion}).

\section{Numerical Results}

Using the established proton and pion LFWFs~\cite{Lan:2021wok,Xu:2021wwj}, we compute the overlap form of the matrix elements in Eqs.~(\ref{eq:amplitudes_proton}) and (\ref{eq:amplitudes_pion}).
For the proton, the four independent twist-3 correlation functions are isolated by applying the projections in Eqs.~(\ref{eq:proj_1})-(\ref{eq:proj_4}).
For the pion, the corresponding function is obtained via the trace projection $T_F^{(\sigma)} = \frac{1}{2M} \hat{M}^{1,[\sigma^{+1}]}_F$.
Our numerical framework naturally covers the hard-pole region where the gluon momentum fraction is non-zero.
However, a direct calculation of the SGP limit ($x_g \to 0$, implying $x_1 = x_2$) is not accessible within the current BLFQ framework due to the absence of zero modes in the longitudinal basis (the plane-wave basis).
Consequently, we estimate the functions in the SGP limit by linearly extrapolating the results obtained in the hard-pole region.

\begin{figure}
  \centering
    \includegraphics[width=0.99\columnwidth]{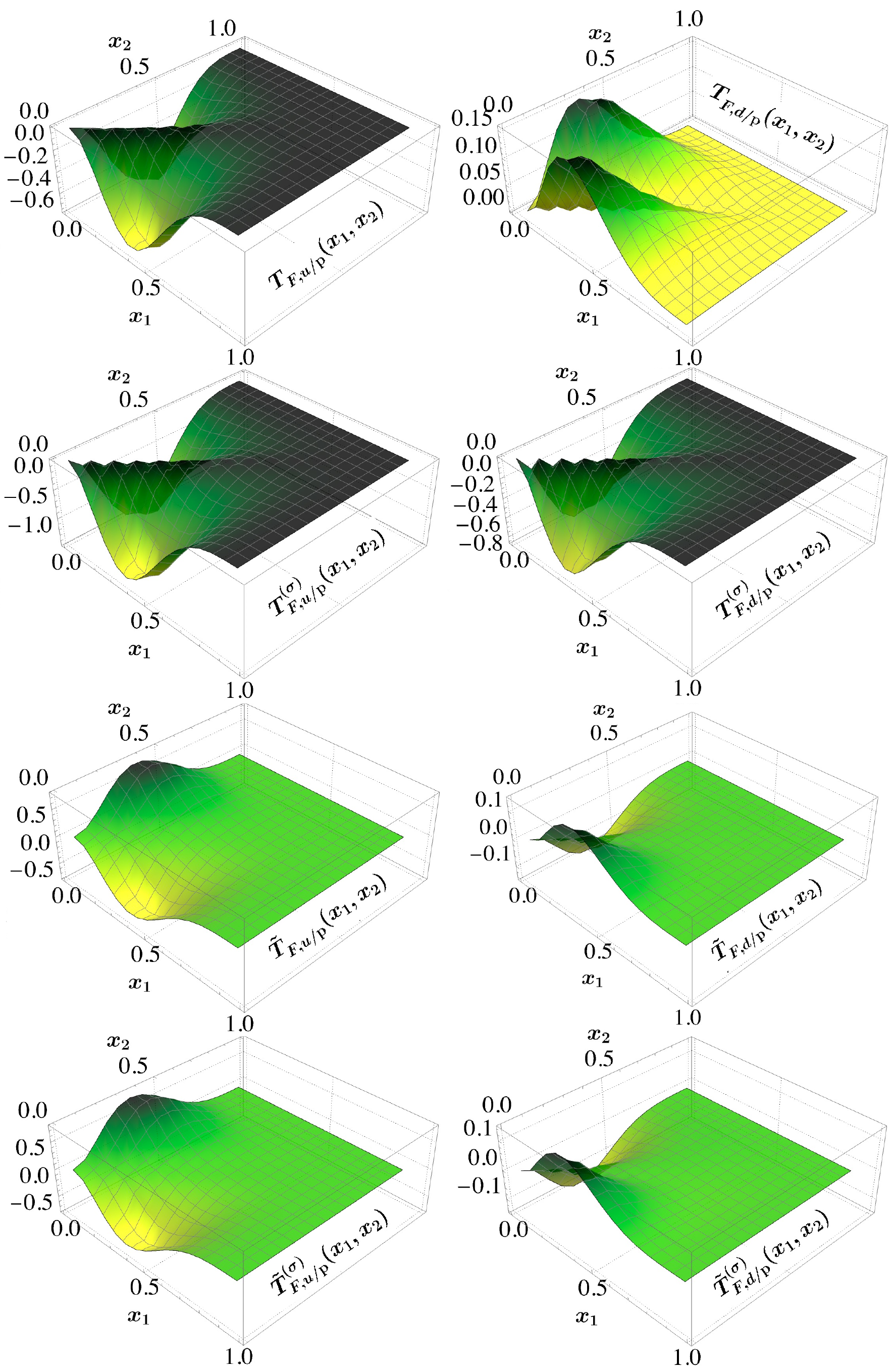}
      \caption{
        Three-dimensional plots of the proton twist-3 functions at the initial scale for the $u$ quark (left column) and the $d$ quark (right column).
        }
      \label{fig:proton3D}
\end{figure}
\begin{figure}
  \centering
        \includegraphics[width=0.99\columnwidth]{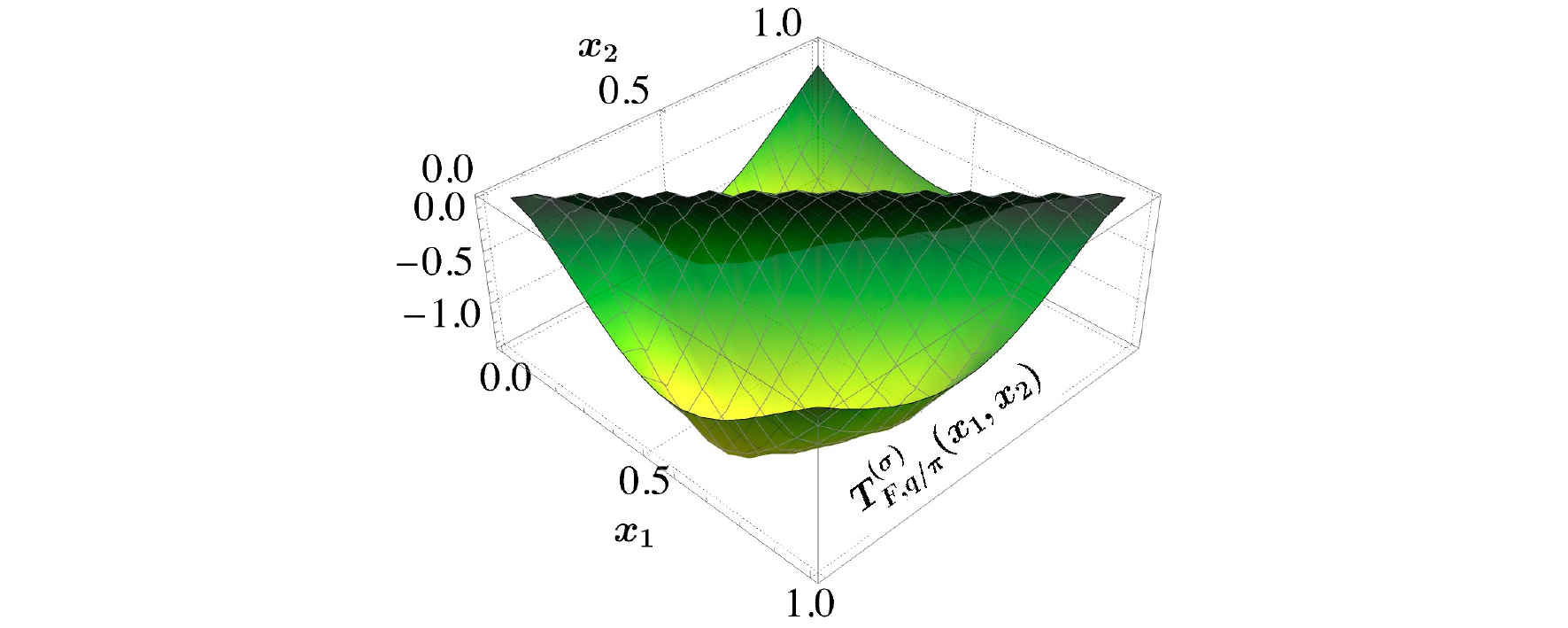}
    \caption{
      Three-dimensional plot of the pion twist-3 function $T_F^{(\sigma)}$ at the initial scale.
    }
    \label{fig:pion3D}
\end{figure}

The resulting twist-3 distributions at the initial scale for the $u$ and $d$ quarks in the proton are presented in Fig.~\ref{fig:proton3D}.
The first two rows correspond to $T_F(x_1,x_2)$ and $T_F^{(\sigma)}(x_1,x_2)$, which are symmetric under $x_1 \leftrightarrow x_2$, while the last two rows display the antisymmetric functions $\tilde{T}_F(x_1,x_2)$ and $\tilde{T}_F^{(\sigma)}(x_1,x_2)$.
There are three notable features emerging from the 3D distributions.
First, the magnitudes of the $u$-quark distributions consistently exceed those of the $d$-quark. 
This enhancement could be due to the proton's valence structure, with twice as many $u$ quarks as $d$ quarks, and the specific spin-flavor structure.
Second, regarding the signs, the ETQS function $T_F(x_1,x_2)$ and the associated functions $\tilde{T}_F(x_1,x_2)$ and $\tilde{T}_F^{(\sigma)}(x_1,x_2)$ exhibit opposite signs for $u$ and $d$ quarks, whereas $T_F^{(\sigma)}(x_1,x_2)$ shares the same sign for both flavors.
Third, for a given flavor, the antisymmetric functions $\tilde{T}_{F,q/\mathrm{p}}(x_1,x_2)$ and $\tilde{T}_{F,q/\mathrm{p}}^{(\sigma)}(x_1,x_2)$ are nearly identical.

Figure~\ref{fig:pion3D} presents the resulting pion twist-3 distribution at the initial scale.
As a spinless meson, the pion only has one spin-independent twist-3 function, $T_F^{(\sigma)}(x_1,x_2)$, which is symmetric under $x_1 \leftrightarrow x_2$.
Notably, the sign of the pion's $T_F^{(\sigma)}(x_1,x_2)$ matches that of the proton's $T_F^{(\sigma)}(x_1,x_2)$.

\begin{figure}[b]
  \centering
  \includegraphics[width=0.99\columnwidth]{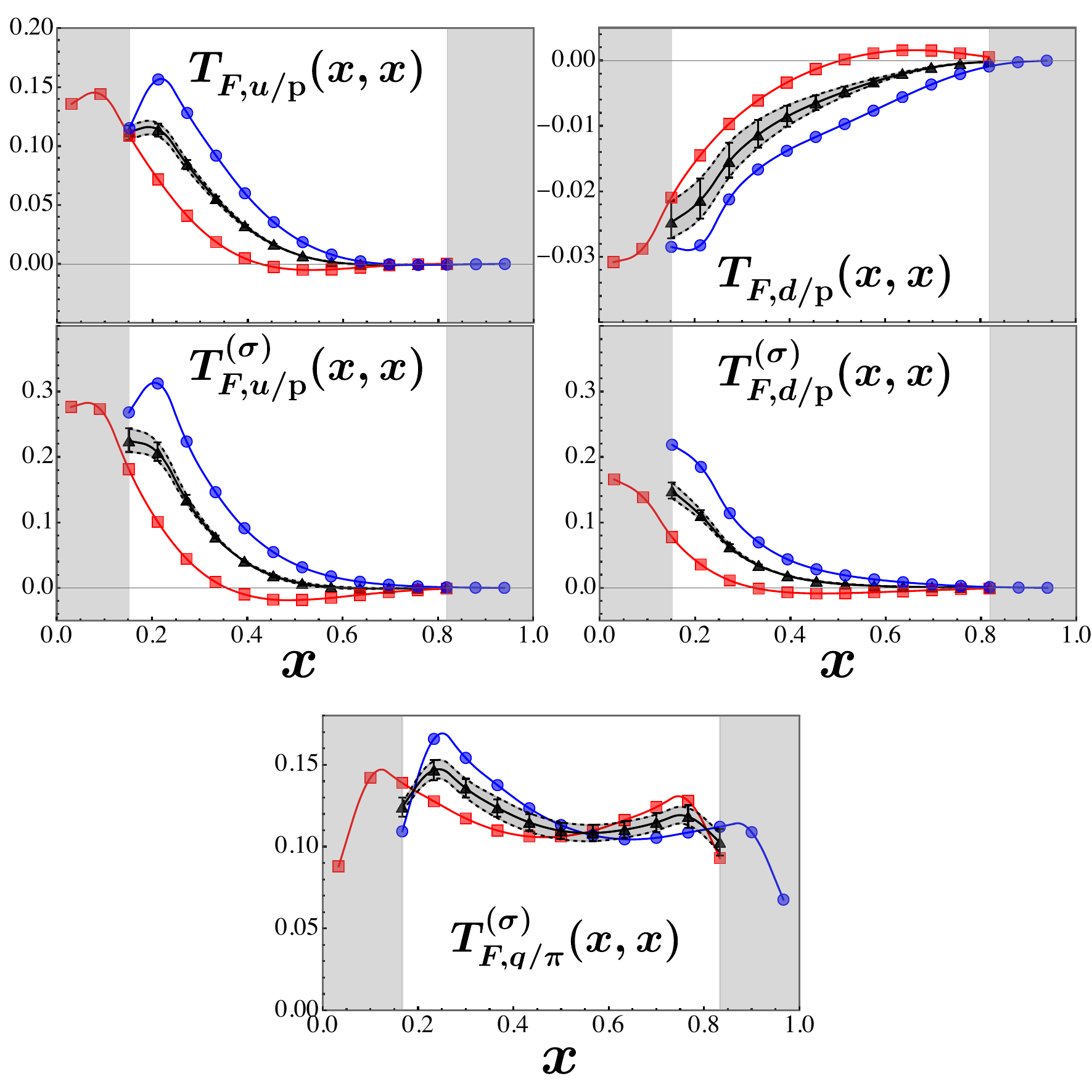}
    \caption{Extrapolated results of the twist-3 functions in the SGP limit for the proton (top two rows) and the pion (bottom row). 
      The black curve represents the average of two linear extrapolations: one from the higher-$x$ region towards the pole (red curve) and one from the lower-$x$ region (blue curve).
      The error bands reflect the combined uncertainty from $\pm10\%$ variations of the initial scale $\mu_0^2$ and the coupling constant $g^2$.
      }
    \label{fig:2D}
\end{figure}

Figure~\ref{fig:2D} represents the twist-3 functions in the SGP limit ($x_1=x_2=x$), obtained via the linear extrapolation.
The red curve is obtained by linearly extrapolating the twist-3 functions in the hard-pole region from large $x$ to small $x$, while the blue curve is obtained by extrapolating from small $x$ to large $x$.
The black curve represents the average of the two extrapolations.
Since linear interpolation requires at least two points, neither two linear interpolation methods can yield a result at one of the boundaries. The corresponding regions are therefore indicated by shaded areas.
The top two rows show the proton results, while the third row shows the pion results.
These SGP results provide a crucial link to experimental observables.
Through model-independent QCD relations, the SGP limit of $T_F(x,x)$ corresponds to the first transverse moment of the Sivers function $f_{1T}^{\perp(1)}(x)$, while $T_F^{(\sigma)}(x,x)$ relates to the Boer-Mulders function $h_{1}^{\perp(1)}(x)$.

Our numerical results at the initial scale are qualitatively consistent with existing experimental extractions and phenomenological calculations~\cite{Gamberg:2007wm,Courtoy:2008dn,Courtoy:2009pc,Barone:2009hw,Gamberg:2009uk,Pasquini:2010af,Lu:2012hh,Pasquini:2014ppa,Wang:2017onm,Li:2019uhj,Tan:2022kgj,Gurjar:2023uho,Lu:2009ip,Longo:2019bih}.
Specifically, the opposite signs of $T_F(x,x)$ for $u$ and $d$ quarks reproduce the well-known experimental observation that the Sivers function has opposite signs for these flavors.
The signs of $T_F^{(\sigma)}(x,x)$ are the same for both quarks, which is consistent with the expectation that the Boer-Mulders functions for $u$ and $d$ quarks have the same sign.
In addition, the sign of $T_F^{(\sigma)}(x,x)$ for the pion is the same as that for the proton, which is consistent with experimental findings indicating that the Boer-Mulders functions for these two hadrons have the same sign.

\end{document}